\documentclass[12pt]{iopart}
\usepackage{graphicx}

\begin{document}

\title[The Buddy Quiz]{The Buddy Quiz: A collaborative assessment and a representation of the scientific enterprise}

\author{Ian M.\ Hoffman}

\address{St.\ Paul's School, 325 Pleasant Street, Concord, New Hampshire, USA 03301}
\ead{ihoffman@sps.edu}
\begin{abstract}
The form and function of a collaborative assessment known as a ``Buddy Quiz'' is presented.
The assessment is conducted in three successive phases over a contiguous 45- to 60-minute class period.
A portion of each Quiz is completed in collaboration with one or two peers and a portion is completed without collaboration.
The Quiz is primarily summative and is also designed to include formative aspects.
The representation in the Quiz of the scientific enterprise as collaborative and individualistic is discussed.
The employment of this instrument in a ninth-grade (age 15 years) conceptual physics course in an independent US secondary school is described and student feedback is presented.
\end{abstract}

\maketitle

\twocolumn
\section{Introduction}

Collaborative assessment is suggested to benefit the individual cognition of students (see, for example, \cite{odo06}).
Peer involvement may take one of two general forms.
(1) Cooperative: assigned social roles in a group whose members' scores depend on the success of the group in executing a well-structured task or (2) collaborative: an ill-structured episode of unconstrained discussion in which individual scores are unrelated to group outcome \cite{coh94,sla95}.
In this paper we describe a written assessment that leverages the latter form of collaboration.
In the assessment described here, an individual student is allowed unconstrained discussion with a peer concerning the material on a written quiz for which the individual student will receive a grade unrelated to the peer.
That is, the students are encouraged, but not required, to use their peers as a resource on a summative assessment; deciding how to employ the peer resource on an assessment is a formative experience.
Following a description by Selker \cite{sel10} of a collaborative assessment technique, we have devised and implemented the ``Buddy Quiz'' described in this paper.

Our exposition of the Buddy Quiz instrument is made in two contexts.
First, the volume of literature is considerable on the subject of peer involvement in educational psychology (see \cite{win09} for a review).
In this light, we examine the Buddy Quiz on cooperative and collaborative grounds in terms of competitive and individualistic environments and intend it to be collaborative and individualistic.
Second, we have designed the Buddy Quiz to support a view that a collaborative and individualistic environment is an authentic representation of the ``scientific enterprise,'' as recommended in policy statements by the National Research Council of the United States \cite{nrc96,nrc01} and the American Association for the Advancement of Science \cite{aaa93}.

In Section~\ref{method} we describe the structure of the assessment instrument and the pedagogical motivations for its various aspects.
In Section~\ref{practical} we describe the implementation of the instrument in a secondary physics classroom, including student feedback.
In Section~\ref{Disc} we explore the Buddy Quiz as a representation of the professional scientific enterprise.

\section{Assessment Method\label{method}}

The Quiz is administered in three phases during a 45- to 60-minute block.
In this section we describe the logistics, goals, and motivations of each phase.
The materials needed are (1) coloured pens for every student, in each of two distinct colours, and (2) a hard copy of the Quiz questions and answer sheet on which each student will write their work for submission.

Buddy Quizzes are summative and the questions are written no differently than for a traditional written exam.
That is, the Quiz is not a series of discussion prompts but rather a few traditional questions with definitive answers.
For example,
\begin{quote}
If two water drops drip successively from a faucet, does the distance between them increase, decrease, or stay the same while they are both falling?
\end{quote}
or
\begin{quote}
If a glass of water and ice cubes is filled to the brim with water, will it overflow as the ice cubes melt?
\end{quote}
or
\begin{quote}
Is the mass of the Moon within one order of magnitude of $10^{25}$~kg?
\end{quote}
are typical questions.
The questions are generally qualitative in nature, although the method described here does not preclude quantitative questions.

\subsection{Individual Phase}

For approximately 15 minutes the students work silently on their own, completing the entire Quiz individually.
For their work during this phase all of the students should use the same coloured pen.
We use traditional, darker colours for this phase such as blue, green, purple, or maroon.
At the end of the time, all of the pens are collected but the Quiz sheet with the individual work is left with the students.

\subsection{Collaborative Phase}

Partners are assigned randomly and the students change their seats in order to be next to their partner.
Pens of a new colour are distributed, the same colour for everyone.
We use lighter, editorial colours for this phase such as pink or orange.
For approximately 15 minutes the students freely discuss the quiz material and amend their individual answers as they see fit, if at all, using the new pen colour.
At the end of the time, all of the pens and all of the Quiz sheets are collected, leaving no materials with the students.

\subsection{Discussion Phase}

For the remaining time, the instructor and students have an open, class-wide discussion reviewing the answers to the questions on the Quiz.

\subsection{Scoring}

For correct answers provided in the first pen colour during the individual phase the students receive full credit.
For correct answers given using the second pen colour during the collaborative phase, students receive half credit (or some other decreased value).
For correct answers in the first colour that are crossed out in favor of an incorrect answer in the second colour, half credit is deducted.

\subsection{Pedagogical Motivations}

As a summative assessment, the availability of full credit during the first phase makes the Buddy Quiz no different than a traditional quiz for the student who is certain of the correct answer.
If certainty is in doubt and the collaborative phase provides the opportunity to gain or to lose points, then the Buddy Quiz is also a formative assessment (see, for example, \cite{bos02}).

Indeed, behaviourally, the students are likely to ask each other ``What did you get for number three?'' at the conclusion of the assessment.
The Buddy Quiz is designed to channel those questions from the close of the individual phase in to the collaborative phase.
The students' adolescent desire to measure themselves against their peers is leveraged.
Also, decision-making in a peer situation is brokered since points are still in jeopardy for amended answers.
Whether or not a student is persuaded by their partner to amend their answer during the collaborative phase is a formative experience that encompasses life skills as well as the course material at hand.

Although the Buddy Quiz offers the potential for formative development, a formal study of student discourse during the collaborative phase is necessary in order to demonstrate formative development (for example, \cite{tao99}).
That is, comments between students during the collaborative phase such as ``I cheated and looked in my notes; the answer is definitely C,'' represent a different formative experience than ``In the limiting case of large distances, the potential approaches zero from the positive side,'' or ``Didn't we see a demo like this?''
The anecdotal student feedback described in Section~\ref{practical} is suggestive of the latter explanation; that the collaborations are formative of physical concepts and the scientific enterprise.

\subsection{Variations}

The Buddy Quiz admits of variations in form and function.
We have designed the instrument for a physics course using input from research in the instruction of natural sciences, but the Quiz seems portable to other subject matter.

Also, changes in the scoring format may stress different aspects.
For example, offering full credit for both the individual phase and the collaborative phase could make the assessment more formative.
Development would be predicated on the students reflecting upon the Quiz phase in which the correct answer was determined.
On the other hand, the Quiz could be made more competitive and summative by keeping track of which partners provide the greatest benefits to their collaborators over many assessments (and rewarding points accordingly).

We have only described examples of qualitative Quiz questions, but quantitative questions of arbitrary difficulty and rote mechanization are also used and may hold significant formative potential.
Again, a study of the discourse is warranted for validation, especially if qualitative and conceptual commentary is expected.

\section{Practical Deployment\label{practical}}

\subsection{Course Description}

In this section we describe the implementation of the Buddy Quiz in two different classes, $X$ and $Y$, of the same Physics First course.\footnote{http://www.aapt.org/Resources/policy/physicsfirst.cfm}
These classes met four times each week for 30 weeks at different times of day in the same room that had five tables at which to work collaboratively.
Section $X$ had 11 students which necessitated four groups of two and one group of three.
Section $Y$ had 14 students which necessitated four groups of three and one group of two.

The grades in the course are comprised of three equal parts: (1) 55-minute exams on which the students work individually, worth twice as much and given half as frequently as Buddy Quizzes, (2) Buddy Quizzes, and (3) individual lab reports on weekly exercises conducted in groups.
Exercises which are not assessed include group discussions on workbook exercises (either from \cite{hew05} or \cite{mcd01}) and class-wide homework review sessions in which preparation is checked for completeness, not veracity.
The grades in the course are not curved.

\subsection{Feedback}

\subsubsection{Testimonials}

Our practice is to have the students write the questions that will appear on the course evaluation questionnaire.
The students each submit questions anonymously and the submissions are compiled into a questionnaire that is administered anonymously.
In this way, both the answers and the questions are useful feedback.
Interestingly, the following question appeared on the questionnaire for Section $Y$: ``Do you like Dr.\ Hoffman's buddy quizzes more than regular quizzes?''
The responses given in Table~1 were collected via anonymous web form after 25 weeks, after approximately ten Quizzes had been administered.

\subsubsection{Teaching Style Inventory}

We also administer the CORD Teaching Style Inventory\footnote{http://www.texascollaborative.org/TSI.htm} to each student, asking each student to answer from the teacher's point of view.
The students were surveyed anonymously on the same day in mid-April as the feedback in Table~\ref{table} was collected.
We present the results here in Figures 1 \& 2 for context on the instructional environment in which the Quizzes are given.

Figure~1 shows that, despite a relatively heavy assessment regimen, the students consider the instruction to favour understanding rather than rote learning.
Similarly, Figure~2 shows that enactive processing is perceived over the symbolic processing required of abstracted lectures.
When interpreting the interaction axis, it is important to note that the text of the TSI describes explicitly cooperative, not collaborative, environments.

\section{Discussion\label{Disc}}

In designing the Buddy Quiz, we have attempted to expose students to the scientific enterprise, as recommended by \cite{aaa93,nrc96,nrc01}.
Thus, our design employed the answers to the following questions:
Is the scientific enterprise collaborative or cooperative?
Is the scientific enterprise competitive or individualistic?
We use the standard definitions of these terms (for example, \cite{coh94,dor97}):
\begin{itemize}
\item[-]{{\bf collaborative:} peer interaction is freely available, encouraged, but not necessary}
\item[-]{{\bf cooperative:} group members have defined roles in a defined task for which all group members receive the same reward}
\item[-]{{\bf competitive:} negative interdependence in which only the highest performers are rewarded}
\item[-]{{\bf individualistic:} the opportunity to attain a reward is not diminished by the presence of capable peers}
\end{itemize}

Is the scientific enterprise collaborative or cooperative?
We consider it to be collaborative.
Research is not a well-structured task mainly because the outcome is not known.
The authorship of a paper, unlike the membership of a cooperative group, is typically uncertain until publication.
Furthermore, the impact of the paper in the literature (the reward) is not dependent on any of the hallmarks of cooperation: the authors are not responsible for ensuring that all of the coauthors have the same understanding of the material or the same share of the work load.

Is the scientific enterprise competitive or individualistic?
We consider it to be individualistic.
A classic competitive environment is one in which test scores are curved; the few ``curve breakers'' at the top will be rewarded while all other peers' rewards suffer because of the presence of the high achievers.
In scientific research, there is no {\it a priori} limit to the number of papers that can be published, no way for one researcher to preclude the success of another when empirical evidence is the only source of merit.
Indeed, consider the free flow of ideas and suggestions at academic colloquia.\cite{ver93,wil06}

Similarly, consider the ubiquitous analogy of swimming races for cooperative and competitive environments.
A four-person relay race is cooperative in that all members must perform in order for a reward to be possible.
A one-person race is competitive in that the only way to define a winner is to define everyone else as losers.
Races are won by the one person or team that subjectively beats the rest of the field; one need not perform well, merely better than everyone else.
In contrast, scientific research papers are published only when the content rises above a level of objective merit, independent of others in the field.

Do peers in science exhibit competition in the vernacular sense?
Yes, of course.
Do politics end up favoring some peers over others for non-objective reasons?
Yes, of course.\cite{etz00}
For example, one important exception to a collaborative and individualistic scientific enterprise occurs in a race to a fundamental (and expensive) discovery: rather than exhibiting the incremental improvement that typifies science, cooperation governs the enterprise with international research agencies authoring papers jointly, as in the field of experimental particle physics.
However, for the most part, the scientific enterprise is not a competitive environment in the pedagogical sense in which peers are hazardous adversaries, rather it is individualistic in which peers are a resource.

Contrary to our current model, others have modelled professional science as a cooperative interplay among non-scientific institutions.\cite{gib94,etz00,tet02,lun10}
Indeed, a contract for ``basic research'' between a government agency and a private corporation is cooperative in that the legal contract is necessarily a well-structured task with well-defined roles and a result determined in advance.
We do not claim that the Buddy Quiz reproduces the societal and political dynamics of these cooperative models, in part because the cooperation in these contracts is among agencies, not the individual scientists.
With the Buddy Quiz, we are seeking to emulate the pure scientific enterprise exemplified by academic research in fields with few ties to industry (for example, astronomy).\cite{mey98}
In the terms of Collins and Evans \cite{col02}, the students are expected to possess interactional expertise and are assessed on their level of contributory expertise.

The collaborations in a Buddy Quiz are limited to small groups in the interest of time and in order to attain and focus the conceptual benefits of the interaction for each student.
Although the scientific enterprise might be better emulated by letting all students mingle freely in search of a consensus answer to each Quiz question, in such an environment we feel that student retention and conceptualization would likely suffer without considerable remedial incentivization atypical of the scientific enterprise.
Rarely would a research discussion be restructured in order to accommodate the least capable and least engaged member; a research group can expel incompetent members but not a physics class its students.

In a Buddy Quiz, each student will submit their own ``paper'' with their own name as first author and they will individually and objectively succeed or fail on decisions they made concerning use of internal and external resources.
The attainment of the correct answer is tantamount, whether it comes after revision is secondary.
The format is overtly summative and individualistic.
Judging from the narrative feedback in Table~\ref{table}, the students perceive the assessment as such and overwhelmingly welcome it.
In addition, formative development is apparent in students who lament losing points after nullifying correct answers during the collaborative phase.

\section{Conclusions}

The ``Buddy Quiz'' is a collaborative assessment that uses standard test questions, two colours of pen for written student work, and sufficient time to allow for peer collaboration and instructor-led discussion.
The design of the Quiz is intended to include both the social-behavioural and the cognitive benefits of peer learning.
The techniques and content of the Buddy Quiz are consistent with recent studies of instruction in the natural sciences and with recommendations from the scientific academy.

\ack

We gratefully acknowledge support from the Lovejoy Science Fund of St.\ Paul's School.
We wish to thank the faculty and students of the St.\ Paul's School Science Division for their support, especially the two sections of Physics First in the Payson Dungeon who are described in this study.

\onecolumn
\section*{References}

\begin{table}[p!]
\caption{\label{table}Student feedback from Section $Y$}
\begin{tabular*}{\textwidth}{l}
\br
Responses to: Do you like Dr.\ Hoffman's buddy quizzes more than regular quizzes? \\
\mr
\begin{minipage}[b]{0.9\textwidth}
\noindent
\begin{itemize}
\item{yes!}
\item{Yes because you can see the thought process of other students}
\item{Yes i do like that they are buddy quiz because i feel that if it were just me taking the quiz i wouldnt learn as much. when i get paired up with a partner they explain it to me and help me understand waht i did wrong. i also like it because i can get points back with a buddy that i wouldnt be able to do if i was alone.}
\item{Yes}
\item{yes, much more}
\item{personally, i think buddy quizzes are more hurtful to my grade because working with a buddy is not helpful to me.}
\item{Yes}
\item{I have no preference}
\item{Yes. The ability to have another person to bounce ideas off of is quite helpful in the learning experience as we can take it from more than one perspective.}
\item{I like them more sometimes but it isn't always fair if you are placed with a weaker partner when others are getting much stronger partners.}
\item{Yes it gives us a better chance to do well and learn.}
\item{yes !}
\item{Yes, they allow you to correct mistakes and still get some credit.}
\item{yes}
\end{itemize}
\end{minipage} \\
\br
\end{tabular*}
\end{table}

\begin{figure}[p]
\centering
\includegraphics[scale=0.75]{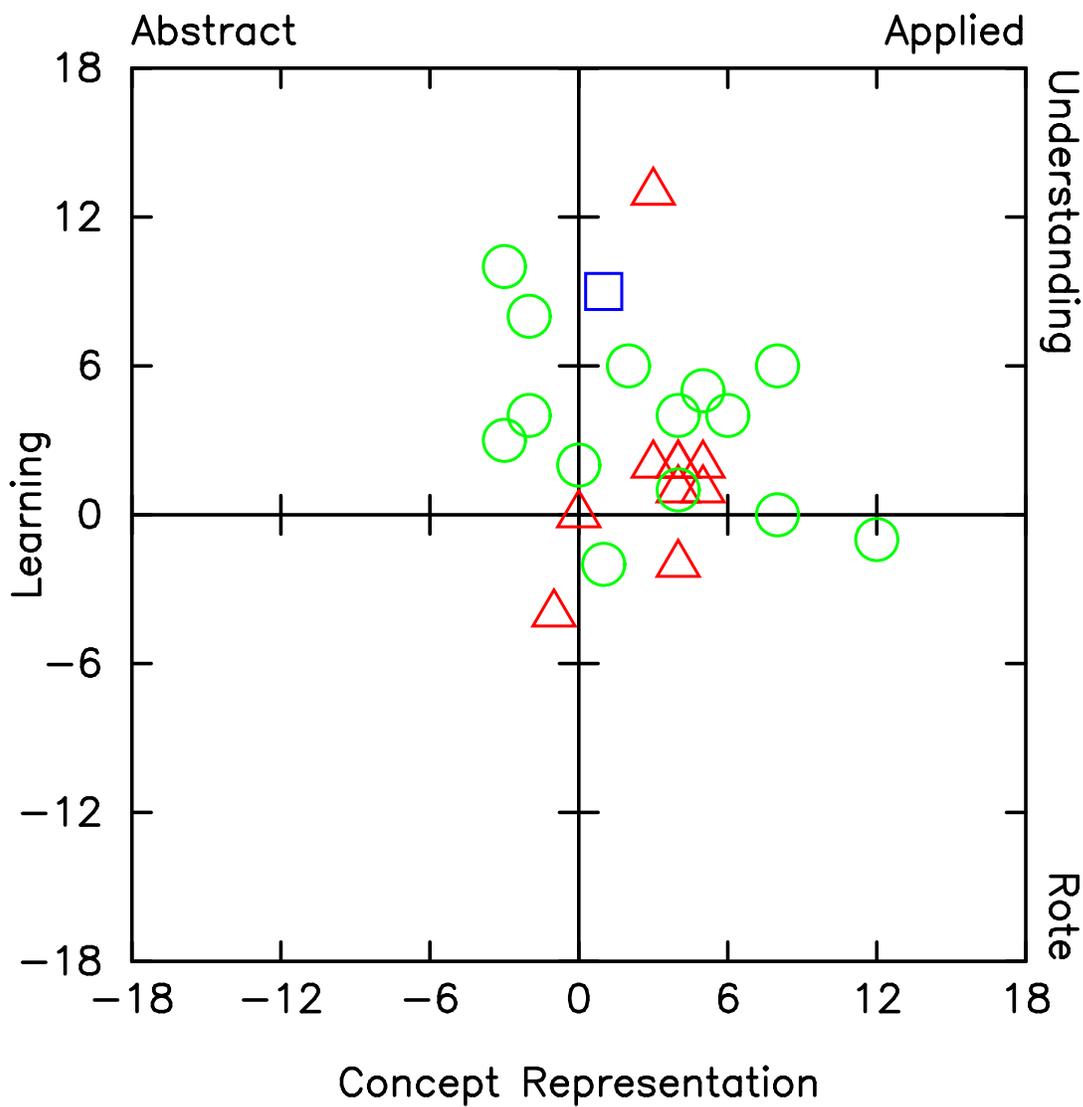}
\caption{
Teaching Goals Matrix.
Student responses to the Teaching Style Inventory described in Section~\ref{practical} for two classes.
The triangles represent Section $X$ $(N=10)$, the circles represent Section $Y$ $(N=14)$, and the square represents the instructor (the author of this paper).
\label{f1}}
\end{figure}

\begin{figure}[p]
\centering
\includegraphics[scale=0.75]{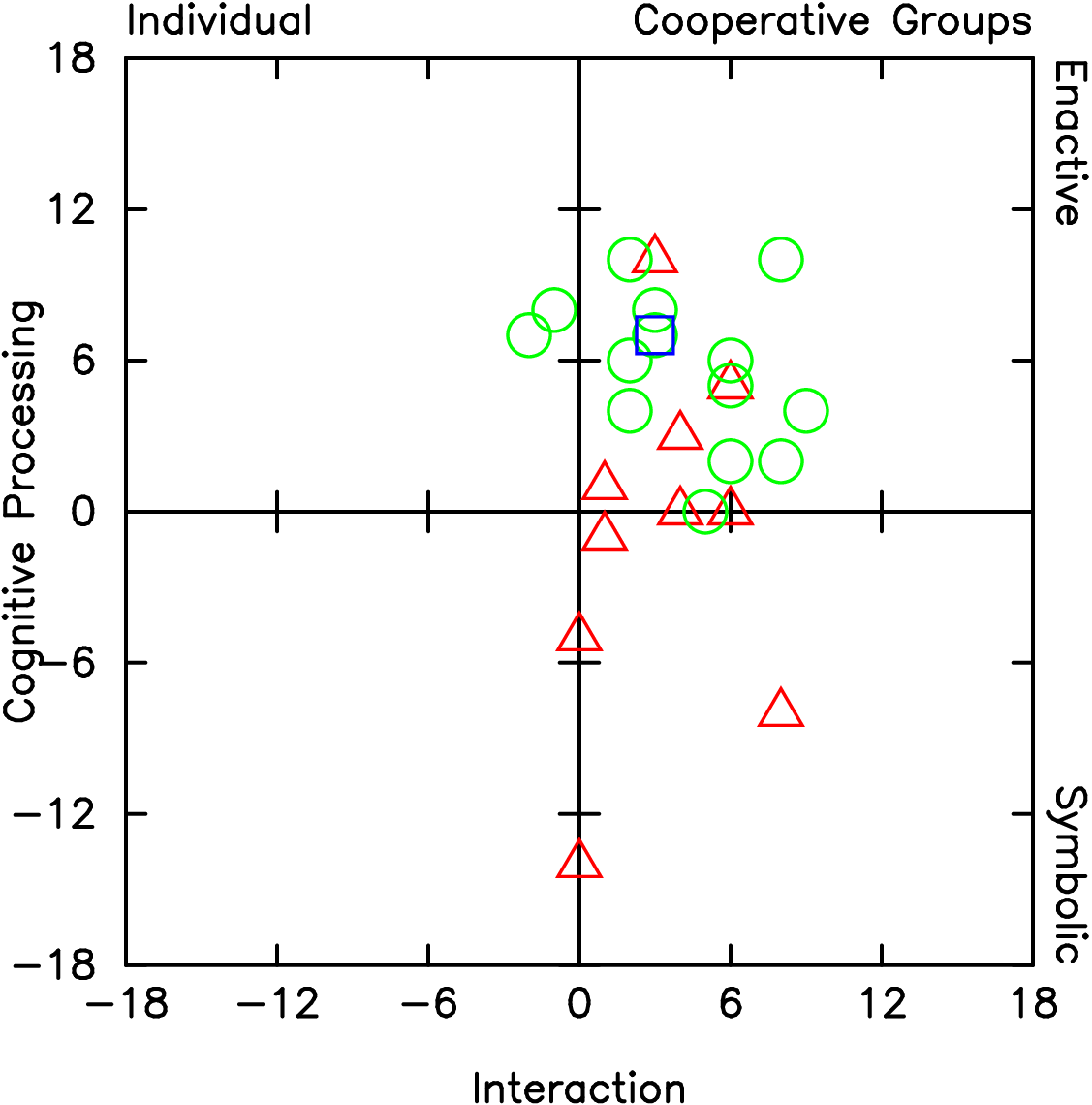}
\caption{
Teaching Methods Matrix.
Student responses to the Teaching Style Inventory described in Section~\ref{practical} for two classes.
The triangles represent Section $X$ $(N=10)$, the circles represent Section $Y$ $(N=14)$, and the square represents the instructor (the author of this paper).
\label{f2}}
\end{figure}

\end{document}